\documentclass{cargese}\usepackage{graphicx}

\let\footnote\savefootnote
\let\footnotetext\savefootnotetext
\let\footnoterule\savefootnoterule
\normallatexbib

\def\be{\begin{equation}}
\def\ee{\end{equation}}

\begin{document}
\articletitle[Towards A  Topological ${\cal G}_2$ string]
{Towards A  Topological ${\cal G}_2$ string\footnote{To appear in the proceedings of Cargese 2004 summer school.}}


\author{Jan de Boer, Asad Naqvi}
\affil{Instituut voor Theoretische Fysica\\
Valckenierstraat 65, 1018XE Amsterdam, The Netherlands }
\email{jdeboer,anaqvi@science.uva.nl }

\author{Assaf Shomer}
\affil{Santa Cruz Institute for Particle Physics\\ 1156 High Street, Santa Cruz, 95064 CA, USA }
\email{shomer@scipp.ucsc.edu}
\vspace{0.5cm}
\noindent hep-th/0502140 \newline
ITFA-2005-03, SCIPP 04/22

\begin{abstract}
We define new topological theories related to sigma models whose target space is a 7 dimensional manifold of $G_2$ holonomy.
We show how to define the topological twist and identify the BRST operator and the physical states. Correlation functions
at genus zero are computed and related to Hitchin's topological action for three-forms. We conjecture that one can
extend this definition to all genus and construct a seven-dimensional topological string theory. In contrast to the
four-dimensional case, it does not seem to compute terms in the low-energy effective action in three dimensions.

\end{abstract}


\section{Introduction}

Topological strings on Calabi-Yau manifolds describe certain
solvable sectors of superstrings and as such provide simplified
toy models of string theory. There are two inequivalent ways to
twist the Calabi-Yau sigma model. This yields topological theories 
known as the A-model and the
B-model, which at first sight depend on different degrees of
freedom: the A-model apparently only involves the K\"ahler moduli
and the B-model only the complex moduli. However, this changes
once branes are included, and it has been conjectured that there
is a version of S-duality which maps the A-model to the B-model
\cite{sdual}. Subsequently, several authors found evidence for the
existence of seven and/or eight dimensional theories that unify
and extend the A and B-model
\cite{gerasimov,vafa,nekrasov,grassi,sinkovics}. This was one of
our original motivations to take a closer look at string theory on
seven-dimensional manifolds of $G_2$ holonomy, and to see whether
it allows for a topological twist, though we were motivated by
other issues as well, such as applications to M-theory
compactifications on $G_2$-manifolds, and as a possible tool to
improve our understanding of the relation between supersymmetric
gauge theories in three and four dimensions.

The outline of this note is as follows. We will first review
sigma-models on target spaces of $G_2$ holonomy, and the structure of chiral the algebra of these theories. The latter is a non-linear
extension of the $c=\frac{21}{2}$ ${\cal N}=1$ superconformal
algebra that contains an ${\cal N}=1$ subalgebra with
$c=\frac{7}{10}$. This describes a minimal model, the tricritical
Ising model, which  plays a crucial role in the twisting. We
then go on to describe the twisting, the BRST operator, the
physical states, and we end with a discussion of topological $G_2$
strings. Here we briefly summarize our findings. A more detailed
discussion will appear elsewhere \cite{toappear}.

There is extensive literature about string theory and
M-theory compactified on $G_2$ manifolds. The first detailed
study of the world-sheet formulation of strings on $G_2$ manifolds
appeared in \cite{sv}.
The world-sheet chiral algebra was studied in some detail in
\cite{sv,blumenhagen,figueroa,noyvert}. For more
about type II strings on $G_2$ manifolds and their mirror
symmetry, see e.g.
\cite{9604133,9707186,0108091,0110302,0111012,0111048,0203272,0204213,0301164,0401125}.
A review of M-theory on $G_2$ manifolds with many references can
be found in \cite{0409191}.

\section{$G_2$ sigma models}

We start from an $\mathcal{N}=(1,1)$ sigma model describing $d$
chiral superfields $X^\mu=\phi^\mu(z) + \theta \psi^\mu(z)$
\be
S= \int d^2z~ d^2 \theta ~(G_{\mu \nu} +B_{\mu \nu}) D_\theta \mathbf{X}^\mu D_{\bar{\theta}} \mathbf{X}^\nu.
\ee
The super stress-energy tensor is given by $T(z,\theta)=G(z)+
\theta T(z) = -{1\over 2}  G_{\mu \nu} D_\theta X^\mu \partial_z
X^\nu$. This $\mathcal{N}=(1,1)$ sigma model can be formulated on
an arbitrary target space. However, the target space theory will
have some supersymmetry only when the manifold has special holonomy.
This condition ensures the existence of covariantly constant spinors which
are used to construct supercharges. The existence of
a covariantly constant spinor on the manifold also implies the
existence of covariantly constant \emph{p-forms} given by
\be\label{peeform}
\phi_{(p)}=\epsilon^T \Gamma_{i_1 \dots i_p} \epsilon ~dx^{i_1}  \wedge \cdots \wedge dx^{i_p}.
\ee
This formal expression may be identically zero. The details of the
target space manifold dictate which p-forms are actually present.
If the manifold has special holonomy $H\subset SO(d)$, the
non-vanishing forms (\ref{peeform}) are precisely the forms that
transform trivially under $H$.

The existence of such covariantly constant p-forms on the target
space manifold implies the existence of extra elements in the
chiral algebra. For example, given a covariantly constant $p$
form, $\phi_{(p)}=\phi_{i_1 \cdots  i_p} dx^{i_1} \wedge \cdots
\wedge dx^{i_p} $ satisfying $\nabla \phi_{i_1 \cdots  i_p}=0$, we
can construct a holomorphic superfield current given by
\[
J_{(p)}(z,\theta)=\phi_{i_1 \cdots  i_p}  D_\theta X^{i_1} \cdots D_{\theta} X^{i_p},
\]
which satisfies $D_{\bar{\theta}} J_{(p)}=0$ on shell. In
components, this implies the existence of a dimension ${p \over
2}$ and a dimension ${p+1 \over 2}$ current. For example, on a
Kahler manifold, the existence of a covariantly constant Kahler
two form $\omega= g_{i\bar{j}}(d \phi^i \wedge
d\phi^{\bar{j}}-d\phi^{\bar{j}} \wedge d\phi^i) $ implies the
existence of a dimension 1 current $J=g_{i\bar{j}} \psi^i
\psi^{\bar{j}}$ and a dimension ${3 \over 2}$ current
$G'(z)=g_{i\bar{j}}(\psi^i \partial_z
\phi^{\bar{j}}-\psi^{\bar{j}} \partial_z \phi^i)$, which add to
the $(1,1)$ superconformal currents $G(z)$ and $T(z)$ to give a
$(2,2)$ superconformal algebra.

A generic seven dimensional Riemannian manifold has $SO(7)$
holonomy. A $G_2 $ manifold has holonomy which sits in a $G_2$
subgroup of $SO(7)$. Under this embedding, the eight dimensional
spinor representation ${\bf 8}$ of $SO(7)$ decomposes into a
$\mathbf{7}$ and a singlet of $G_2$, and the latter corresponds to the covariantly constant spinor. The $p$-form (\ref{peeform}) is
non-trivial only when $p=~3,~ 4$. In other words, there is a
covariantly constant 3-form $\phi^{(3)} = \phi_{ijk}^{(3)} dx^i
\wedge dx^j \wedge dx^k $. The hodge dual 4-form is then also
automatically covariantly constant.

By the above discussion, the 3-form implies the existence of a
superfield current $J_{(3)}(z,\theta)=\phi_{ijk}^{(3)} D_\theta
X^i D_\theta X^j D_\theta X^k \equiv  \Phi +\theta K$. Explicitly,
$\Phi$ is a  dimension ${3 \over 2}$ current
$\Phi=\phi_{ijk}^{(3)} \psi^i \psi^j \psi^k$ and $K$ is its
dimension 2 superpartner $K=\phi_{ijk}^{(3)} \psi^i \psi^k\partial
\phi^k$. Similarly, the 4-from implies the existence of a
dimension 2 current $X$ and its dimension ${5\over 2}$
superpartner $M$. The chiral algebra of $G_2$ sigma models thus
contains 4 extra currents on top of the two $G,T$ that constitute
the ${\cal N}=1$ superconformal algebra. These six generators form
a closed quantum algebra which appears explicitly e.g. in
\cite{blumenhagen,sv,figueroa} (see also \cite{noyvert}).

An important fact, which will be crucial in almost all the
remaining analysis, is that the generators $\Phi$ and $X$ form a
closed sub-algebra: if we define the supercurrent $G_{I}={i \over
\sqrt{15}} \Phi$ and stress-energy tensor $T_I=-{1 \over 5}X$ we
recognize that this is the unique ${\cal N}=1$ super-conformal
algebra of the minimal model with central charge $c={7 \over 10}$
known as the \emph{Tri-critical Ising Model}. This sub-algebra plays a role similar to the $U(1)$
R-symmetry of the $\mathcal{N}=2$ algebra  in compactifications on Calabi-Yau manifolds.

In fact, with respect to the conformal symmetry, the full Virasoro
algebra decomposes in two commuting\footnote{This decomposition only works
for the Virasoro part of the corresponding $\mathcal{N}=1$ algebras. The full $\mathcal{N}=1$ structures  do
not commute. For example the superpartner of $\Phi$ with
respect to the full ${\cal N}=1$ algebra is $K$ whereas its
superpartner with respect to the ${\cal N}=1$ of the tri-critical
Ising model is $X$. } Virasoro algebras: $T=T_I+T_r$ with $T_I(z)
T_r(w)=0.$ This means we can classify conformal primaries by two
quantum numbers, namely its tri-critical Ising model highest
weight and its highest weight with respect to $T_r$: $|{\rm
primary} \rangle = | h_I,h_r \rangle $.

Perhaps it is worth emphasizing the logic here: classically, we
find a conformal algebra with six generators in sigma-models on
manifolds of $G_2$ holonomy. In the quantum theory we expect, in
the absence of anomalies other than the conformal anomaly, to find
a quantum version of this classical algebra. In \cite{blumenhagen}
all quantum extensions were analyzed, and a two-parameter family
of quantum algebras was found. Requiring that the quantum algebra
has the right central charge (necessary to have a critical string
theory) and that it contains the tricritical Ising model
(necessary for space-time supersymmetry) fixes the two-parameters.
This motivates why this is the appropriate definition for string
theory on $G_2$ manifolds.

\section{Tri-Critical Ising Model} \label{tim}

Unitary minimal models are labeled by a positive integer
$p=2,3,\dots$ and occur only on the ``discrete series" at central
charges $c=1-\frac{6}{p(p+1)}$. The Tri-Critical Ising Model is
the second member ($p=4$) which has central charge
$c=\frac{7}{10}$. It is at the same time also a minimal model for
the $\mathcal{N}=1$ superconformal algebra.

The conformal primaries of unitary minimal models are labeled by
two integers $1\leq n'\leq p$ and $1\leq n<p$. The weights in this
range are arranged into a ``\emph{Kac table}".  The conformal
weight of the primary $\Phi_{n'n}$ is
$h_{n'n}=\frac{[pn'-(p+1)n]^2-1}{4p(p+1)}.$ In the Tri-critical
Ising model $(p=4)$ there are 6 primaries of weights
$0,\frac{1}{10},\frac{6}{10},\frac{3}{2},\frac{7}{16},\frac{3}{80}$.
Below we write the Kac table for the tricritical Ising model.
Beside the Identity operator $(h=0)$ and the ${\cal N}=1$
supercurrent $(h=\frac{3}{2})$ the NS sector (first and third
columns) contains a primary of weight $h=\frac{1}{10}$ and its
${\cal N}=1$ superpartner $(h=\frac{6}{10})$. The primaries of
weight $\frac{7}{16},\frac{3}{80}$ are in the Ramond sector
(middle column).

\begin{center}
 \begin{tabular}{|c|c|c|c|}
   \hline

 $n' \setminus n $ & $ 1$ & $ 2$ & $ 3$ \\
     \hline
    1 & $\mathbf{0}$ & $\mathbf{\frac{7}{16}}$\  & $\frac{3}{2}$\\
  \hline
    2 & $\mathbf{\frac{1}{10}\ }$ & $ \mathbf{\frac{3}{80}\ }$ &
 $\frac{6}{10}\ $ \\   \hline
   3 & $ \mathbf{\frac{6}{10}\ }$ & $\frac{3}{80}\ $ &
 $\frac{1}{10}\ $ \\  \hline
   4 & $ \mathbf{\frac{3}{2}\ }$ & $\frac{7}{16}\ $ & 0 \\
 \hline
   \end{tabular}
\end{center}

The Hilbert space of the theory decomposes in a similar way,
${\cal H}=\oplus_{n,n'} {\cal H}_{n',n}$. A central theme in this
work relies on the fact that since the primaries $\Phi_{n'n}$ form
a closed algebra under the OPE they can be decomposed into
\emph{conformal blocks}\ which connect two Hilbert spaces.
Conformal blocks are denoted by $\Phi_{n',n,m'm}^{l',l}$ which
describes the restriction of $\Phi_{n',n}$ to a map that only acts
from $\mathcal{H}_{m',m}$ to $\mathcal{H}_{l',l}$.

An illustrative example, which will prove crucial in what follows,
is the block structure of the primary $\Phi_{2,1}$ of weight
$1/10$. General arguments show that the fusion rule of this field
with any other primary $\Phi_{n'n}$ is $\phi_{(2,1)}\quad\times
\quad\phi_{(n',n)}=\phi_{(n'-1,n)}\quad +\quad\phi_{(n'+1,n)}.$
The only non-vanishing conformal blocks in the decomposition of
$\Phi_{2,1}$ are those that connect a primary with the primary
right above it and the primary right below in the Kac table,
namely\footnote{Note the confusing notation where \emph{down} the
Kac table means \emph{larger} n' and vice-versa.},
$\phi_{2,1,n',n}^{n'-1,n}$ and $\phi_{2,1,n',n}^{n'+1,n}$. This
can be summarized formally by defining the following
decomposition\footnote{We stress that this decomposition is
special to the field $\Phi_{2,1}$ and does not necessarily hold
for other primaries which may contain other blocks.}
\be\label{updown}
\Phi_{2,1}=\Phi_{2,1}^{\downarrow} \oplus \Phi_{2,1}^{\uparrow}.
\ee
Similarly, the fusion rule of the Ramond field $\Phi_{1,2}$ with
any primary is $\phi_{(1,2)}\quad\times
\quad\phi_{(n',n)}=\phi_{(n,n-1)}\quad+\quad\phi_{(n',n+1)}$
showing that it is composed of two blocks, which we denote as
follows $\Phi_{1,2}=\Phi_{1,2}^{-} \oplus \Phi_{1,2}^{+}.$
Conformal blocks transforms under conformal transformations
exactly like the primary field they reside in but are usually not
 single-valued functions of $z(\bar{z})$.

\subsection{Chiral Primary States}\label{ssam}

The chiral-algebra associated with manifolds of $G_2$
holonomy\footnote{We loosely refer to it as ``the $G_2$ algebra"
but it should not be confused with the Lie algebra of the group
$G_2$.} allows us to draw several conclusions about the possible
spectrum of such theories. It is useful to decompose the
generators of the chiral algebra in terms of primaries of the
tri-critical Ising model and primaries of the
remainder. The commutation
relations of the $G_2$ algebra imply that the some of the generators
of the chiral algebra decompose as \cite{sv}:
$G(z)=\Phi_{2,1}\otimes\psi_{\frac{14}{10}}\quad,\quad
K(z)=\Phi_{3,1}\otimes\psi_{\frac{14}{10}}$ and
$M(z)=a\Phi_{2,1}\otimes\chi_{\frac{24}{10}}+b[X_{-1},\Phi_{2,1}]\otimes\psi_{\frac{14}{10}},$
with $\psi,\chi$ primaries of the indicated weights in the $T_r$
CFT and $a,b$ constants.

Ramond ground states of the full $c={21\over 2}$ SCFT are of the
form $|{7\over 16},0\rangle$ and $|{3\over 80},{2\over 5}\rangle$.
The existence of the $|{7\over 16},0\rangle$ state living just
inside the tricritical Ising model plays a
crucial role in the topological twist. Coupling left and right
movers, the only possible RR ground states compatible with the
$G_2$ chiral algebra\footnote{Otherwise the spectrum will contain
a 1-form which will enhance the chiral algebra. Geometrically this
is equivalent to demanding that $b_1=0$.} are a single $|{7\over
16},0\rangle_L\otimes\ |{7\over 16},0\rangle_R$ ground state and a
certain number of states of the form $|{3\over 80},{2\over
5}\rangle_L\otimes\ |{3\over 80},{2\over 5}\rangle_R$. By studying
operator product expansions of the RR ground states we get the
following ``special" NSNS states $|0,0\rangle_L\otimes\
|0,0\rangle_R,\ |{1\over 10},{2\over 5}\rangle_L\otimes\ |{1\over
10},{2\over 5}\rangle_R,\ |{6\over 10},{2\over 5}\rangle_L\otimes\
|{6\over 10},{2\over 5}\rangle_R$ and $|{3\over
2},0\rangle_L\otimes\ |{3\over 2},0\rangle_R$ corresponding
 to the 4 NS primaries $\Phi_{n',1}$ with $n'=1,2,3,4$ in the tri-critical Ising model.
Note that for these four states there is a linear relation between
the Kac label $n'$ of the tri-critical Ising model part and the
total conformal weight $h_{total}={n'-1\over 2}$. In fact, it can
be shown that, similar to the BPS bound in the $\mathcal{N}=2$
case, primaries of the $G_2$ chiral algebra satisfy a (non-linear)
bound of the form
\be \label{bound}
h_I + h_r \geq
\frac{1+\sqrt{1+80 h_I}}{8} .
 \ee
which is precisely saturated for the four NS states listed above.
We  will therefore refer to those states as ``chiral primary"
states. Just like in the case of Calabi-Yau, the ${7\over 16}$
field maps Ramond ground states to NS chiral primaries and is thus
an analogue of the ``spectral flow" operators in Calabi-Yau.

\section{Topological Twist}

To construct a topologically twisted CFT we usually proceed in two steps. First
we define a new stress-energy tensor, which changes the quantum numbers of the fields
and operators of the theory under Lorentz transformations. Secondly, we identify a
nilpotent scalar operator, usually constructed out of the supersymmetry generators of the
original theory, which we declare to be the BRST operator. Often this BRST operator
can be obtained in the usual way by gauge fixing a suitable symmetry. If the new
stress tensor is exact with respect to the BRST operator, observables (which are elements
of the BRST cohomology) are metric independent and the theory is called topological.
In particular, the twisted stress tensor should have a vanishing central charge.

In practice \cite{kodira,antoniadis}, for the ${\cal N}=2$
theories, an n-point correlator on the sphere in the twisted theory
can conveniently be defined\footnote{Up to proper normalization.}
as a correlator in the \emph{untwisted} theory of the same n
operators plus two insertions of a spin-field, related to the
space-time supersymmetry charge, that serves to trivialize the
spin bundle. For a Calabi-Yau 3-fold target space there are two
$SU(3)$ invariant spin-fields which are the two spectral flow
operators $\mathcal{U}_{\pm{1\over 2}}$. This discrete choice in the left
and the right moving sectors is
the choice between the $+(-)$ twists \cite{Witten} which results
in the difference between the topological $A/B$ models.

In \cite{sv} a similar expression was written down for sigma models on $G_2$ manifolds,
this time involving the single $G_2$ invariant spin field
which is the unique primary $\Phi_{1,2}$ of weight ${7\over 16}$.
It was proposed that this expression could be a suitable definition of the correlation
functions of a putative
topologically twisted $G_2$ theory.
In other words the twisted amplitudes are defined as\footnote{Up to a coordinate
dependent factor that we omit here for brevity and can be found in \cite{toappear}.}
\be\label{twicorsphere}
\langle V_1(z_1)\dots V_n(z_n)\rangle_{\mathtt{twist}}\equiv
\langle \Sigma(\infty)V_1(z_1)\dots V_n(z_n)\Sigma(0)\rangle_{\mathtt{untwist}}.
\ee
In \cite{sv} further arguments were given, using the Coulomb gas representation
of the minimal model, that there exists a twisted stress tensor with vanishing central
charge. This argument is however problematic, since the twisted stress tensor 
proposed there does not commute with Felder's BRST operators \cite{felder} and therefore
it does define a bona fide operator in the minimal model. In addition, a precise
definition of a BRST operator was lacking.

We will proceed somewhat differently. We will first propose a BRST operator, study
its cohomology, and then use a version of (\ref{twicorsphere}) to compute
correlation functions
of BRST invariant observables. We will then comment on the extension to higher genus
and on the existence of a topologically twisted $G_2$ string.

\section{The BRST Operator}

Our basic idea is that the topological theory for $G_2$ sigma models should be
 formulated not in terms of local operators of the untwisted theory
but in terms of its (non-local)\footnote{It should be
stressed that this splitting into
conformal blocks is non-local
in the simple sense that conformal blocks may be
multi-valued functions of $z(\bar{z})$.} conformal blocks.

By using the decomposition (\ref{updown}) into conformal blocks, we can split any
field whose tri-critical Ising model part contains just the conformal family
$\Phi_{2,1}$ into its \emph{up} and \emph{down} parts. In particular, the ${\cal N}=1$ supercurrent $G(z)$ can be split as
\be\label{gmsplit} G(z)=G^{\downarrow}(z) + G^{\uparrow}(z). \ee

We claim that $G^{\downarrow}$ is the BRST current and
$G^{\uparrow}$ carries many features of an anti-ghost. 

The basic ${\cal N}=1$ relation \be G(z)G(0)=\left(
G^{\downarrow}(z)+G^{\uparrow}(z)\right)\left(
G^{\downarrow}(0)+G^{\uparrow}(0)\right)\sim
\frac{2c/3}{z^3}+\frac{2T(0)}{z} \ee
proves the nilpotency of this BRST current (and of the candidate
anti-ghost) because the RHS contains descendants of the identity operator only
and has trivial fusion rules with the primary fields of the tri-critical Ising model
and so $(G^{\downarrow})^2=(G^{\uparrow})^2=0$.

More formally, denote by $P_{n'}\ $ the projection operator
on the sub-space ${\cal H}_{n'}$ of states whose tri-critical Ising model
part lies within the conformal family of one of the
four NS primaries $\Phi_{n',1}$. The 4 projectors add to the identity
\be
\label{j11} P_1+P_2+P_3+P_4=1,
\ee
because this exhaust the list of possible highest weights in the
NS sector of the tri-critical Ising model\footnote{For simplicity,
we will set $P_{n'}=0$ for $n'\leq 0$ and $n'\geq 5$, so that we
can simply write $\sum_{n'} P_{n'}=1$ instead of (\ref{j11}).}. We
can now define our BRST operator in the NS sector more rigorously
\be\label{BRSTop}
Q = G_{-{1\over 2}}^{\downarrow}\equiv\sum_{n'} P_{n'+1} G_{-{1\over 2}} P_{n'}.
\ee
The nilpotency $Q^2=0$ is easily proved
\be Q^2 = \sum_{n'} P_{n'+2}
G_{-{1\over2}}^2 P_{n'} = \sum_{n'} P_{n'+2} L_{-1} P_{n'}=0,
\ee
where we could replace the intermediate $P_{n'+1}$ by the identity because of
 property \ref{gmsplit}\ and the last equality follows
since $L_{-1}$ maps each ${\cal H}_{n'}$ to itself.

$Q$ does not commute with the local operator
$\mathcal{O}_{\Delta(1),0},\ \mathcal{O}_{\Delta(2),{2\over 5}},$
$\mathcal{O}_{\Delta(3),{2\over 5}}$ and
$\mathcal{O}_{\Delta(4),0}$ corresponding to the chiral states
$|0,0\rangle, |\frac{1}{10},\frac{2}{5}\rangle,
|\frac{6}{10},\frac{2}{5}\rangle$ and $|\frac{3}{2},0\rangle$ (for
brevity we will denote those 4 local operators juts by their
minimal model Kac label $\mathcal{O}_i,\ i=1,2,3,4).\ $ However,
it can be checked that the following blocks,
\be\label{cpbloc}
\mathcal{A}_{n'}=\sum_{m}^{}P_{n'+m-1}\mathcal{O}_{n'}P_{m},
\ee
which pick out the maximal ``down component" of the corresponding
local operator, do commute with $Q$ and are thus in its operator
cohomology. Thus the chiral \emph{operators}\ of the twisted model
are represented in terms of the blocks \ref{cpbloc}\ of the local
operators corresponding to the chiral \emph{states}. Furthermore, it can be
 shown easily that those chiral operator form a ring under
the OPE.

By doing a calculation at large volume, we see that the BRST cohomology has one
operator of type ${\cal O}_1 \otimes {\cal O}_1$, $b_2+b_3$ of type
${\cal O}_2\otimes {\cal O}_2$, $b_4+b_5$ of type
${\cal O}_3 \otimes {\cal O}_3$, and one of type ${\cal O}_4 \otimes {\cal O}_4$.
The total BRST cohomology is thus precisely given by $H^{\ast}(M)$.
Also, one finds that the $b_3$ operators of type ${\cal O}_2\otimes {\cal O}_2$
are precisely the geometric moduli of the $G_2$ target space\footnote{In this work we ignore the $b_2$ moduli corresponding to the $B$-field.}.

In the topological $G_2$ theory, genus zero correlation functions
of chiral primaries between BRST closed states are position
independent. Indeed, the generator of translations on the plane,
namely $L_{-1}$, is BRST exact

\be
\{ G^{\downarrow}_{-{1\over 2}},G^{\uparrow}_{-{1\over
2}}\}=\sum_{n'}P_{n'}G_{-{1\over 2}}\left( P_{n'-1}+
P_{n'+1}\right)G_{-{1\over 2}}
P_{n'}=\sum_{n'}P_{n'}L_{-1}P_{n'}=L_{-1}.
\ee This is a crucial ingredient of topological theories.

Moreover, it can be shown  \cite{sv}\ that the upper components
$\tilde{G}_{-{1\over 2}}|{1\over 10},{2\over 5}\rangle_L\otimes\
G_{-{1\over 2}}|{1\over 10},{2\over 5}\rangle_R$ correspond to
exactly marginal deformations of the CFT preserving the $G_2$
chiral algebra, completely in agreement with the identification of
them as the geometric moduli of the theory. Focusing momentarily
on the left movers, we can show that $[Q,\{ G_{-{1\over
2}},\mathcal{O}_2 \}]=
\partial\mathcal{A}_2$\,\footnote{Recall that ${\cal A}_2$ was defined in
(\ref{cpbloc}).} so that the very same deformation is physical,
namely Q exact, also in the topological theory. Note that the
deformation is given by a conventional operator that does not
involve any projectors. Combining with the right-movers, we find
that the deformations in the action of the topological string are
exactly the same as the deformations of the non-topological string
as is expected because both should exist on an arbitrary manifold
of $G_2$ holonomy.

Most correlation functions at genus zero vanish. The most
interesting one is the three-point function of three operators
$Y={\cal O}_2\otimes {\cal O}_2$. These correspond to geometric
moduli. If we introduce coordinates $t_i$ on the moduli space of $G_2$ metrics,
then we obtain from a large volume calculation
\begin{equation}
\langle Y_i Y_j Y_k \rangle = \int_M d^7 x \sqrt{g} \phi_{abc}
\frac{\partial g^{aa'}}{\partial t_i} \frac{\partial
g^{bb'}}{\partial t_j}\frac{\partial g^{cc'}}{\partial t_k}
\phi_{a'b'c'}.
\end{equation}
One might expect, based on general arguments, that this is the
third derivative of some prepotential if suitable `flat'
coordinates are used. We do not know the precise definition of
flat coordinates for the moduli space of $G_2$ metrics, but if we
take for example $M=T^7$ and take coordinates such that $\phi$ is
linear in them, then we can verify
\begin{equation}
\langle Y_i Y_j Y_k \rangle = -\frac{1}{21}
\frac{\partial^3}{\partial t_i \partial t_j \partial t_k} \int
\phi \wedge \ast \phi.
\end{equation}
The prepotential appearing on the right hand side is exactly the
same as the action functional introduced by Hitchin in
\cite{hitchin,hitchin2}. A similar action was also used as a starting point
for topological M-theory in \cite{vafa} (see also \cite{gerasimov}). This strongly suggests that our
topological $G_2$ field theory is somehow related to topological
M-theory.

\section{Topological $G_2$ Strings}\label{topologicalstrings}

In the case of $N=2$ theories, the computation of correlation
functions at genus zero outlined above can be generalized to
higher genera \cite{kodira,antoniadis}. An n-point correlator on a
genus-g Riemann surface in the twisted theory can be defined as a
correlator in the untwisted theory of the same n operators plus
$(2-2g)$ insertions of the spin-field that is related to the
space-time supersymmetry charge. For a Calabi-Yau 3-fold target
space on a Riemann surface with $g>1$, the meaning of the above
prescription is to insert $2g-2$ of the conjugate spectral flow
operator.

To generalize this to the $G_2$ situation, we would like to have
something similar. However, there is only one $G_2$ invariant
spin-field. This is where the decomposition in conformal blocks
given in section~\ref{tim} is useful: the spin-field $\Phi_{2,1}$
could be decomposed\footnote{In terms of the Coulomb gas
representation, one of these can be represented as an ordinary
vertex operator, the other one involves a screening charge.} in a
block $\Phi_{2,1}^+$ and in a block $\Phi_{2,1}^-$. At genus zero
we needed two insertions of $\Phi_{2,1}^+$, so the natural guess
is that at genus $g$ we need $2g-2$ insertions of $\Phi_{2,1}^-$.
However, this is not the full story. We also need to insert $3g-3$
copies of the anti-ghost and integrate over the moduli space of
Riemann surfaces to properly define a topological string theory.
The anti-ghost is very close to $G^{\uparrow}$, and the fusion
rules of the tri-critical Ising model tell us that there is indeed
a non-vanishing contribution to correlation functions of $2g-2$
$\Phi_{2,1}^-$'s and $3g-3$ $G^{\uparrow}$.

This prescription would therefore work very nicely if we would
have found the right anti-ghost. The candidates we tried so far all
seem to fail in one way or another. One possible conclusion might
be that a twisted stress tensor does not exist and that there is
only a sensible notion of topological $G_2$ sigma models but not
of topological $G_2$ strings. The fact that so far so many
properties of the ${\cal N}=2$ topological theories appeared to
hold also in our $G_2$ model leads us to believe that a sensible
extension to higher genera indeed exists. Identifying the correct
twisted stress tensor remains an open problem. Barring this
important omission the coupling to topological gravity goes pretty
much along the same lines as for the $\mathcal{N}=2$ topological
string (details can be found in \cite{toappear}).

\section{Conclusions}

An important application of topological strings stems from the
realization \cite{kodira,antoniadis,Witten} that its amplitudes
agree with certain amplitudes of the physical superstring. Just
like ${\cal N}=2$ topological strings compute certain F-terms in
four dimensional ${\cal N}=2$ gauge theories, one might wonder
whether the $G_2$ topological string similarly computes F-terms in
three dimensional ${\cal N}=2$ gauge theories.

Since $G_2$ manifolds are Ricci-flat we can consider compactifying
the type II superstring on $R^{1,2}\times\mathcal{N}_7$ where
$\mathcal{N}_7$ is a 7 dimensional manifold of $G_2$ holonomy.
This reduces the supersymmetry down to two real supercharges in
3 dimension from each worldsheet chirality so we end up with a low
energy field theory in 3 dimensions with ${\cal N}=2$
supergravity. By studying amplitudes in some detail we observe
that, except perhaps at genus zero, the amplitudes do involve a
sum over conformal blocks, and the topological $G_2$ string
therefore seems to compute only one of many contributions to an amplitude. This is in contrast to the four dimensional case where the possibility to look only at the (anti)-self dual gravitons allowed to isolate the topological contributions.

Nevertheless, we believe that topological $G_2$ strings are
worthwhile to study. They may possibly provide a good definition
of topological M-theory, and a further study may teach us many
things about non-topological $G_2$ compactifications as well. We leave this, as well as the generalization to $spin(7)$ 
compactifications \cite{spin7}\ and the study of branes in these theories to future work.

\newpage

\begin{acknowledgments}
It is a pleasure to thank Nathan Berkovits, Volker Braun, Lorenzo Cornalba, Robbert Dijkgraaf, Pietro Antonio Grassi, Wolfgang Lerche, Hiroshi Ooguri, Volker Shomerus,  Annamaria Sinkovics, Kostas Skenderis and Cumrun Vafa for useful discussions.
We would also like to
thank the other organizers of the Carg\`ese 2004 summer school.
This work was partly supported by the stichting FOM.

\end{acknowledgments}


\end{document}